\documentclass[12pt,twoside,epsfig,here,graphicx]{article}
\usepackage{geometry}
\usepackage{epsfig}
\usepackage{graphics}
\usepackage{extramarks}

\begin{document}                                        % do not modify
\begin{center}                                          % do not modify
%%%%%%%%%%%%%%%%%%%%%%%%%%%%%%%%%%%%%%%%%%%%%%%%%%%%%%%%%%%%%%%%%%%%%%%%
%% Title of the proceeding replace with your title.
%%
\large{\bf Neutral and charged pion production in the S-matrix approach} \\
%%
%%%%%%%%%%%%%%%%%%%%%%%%%%%%%%%%%%%%%%%%%%%%%%%%%%%%%%%%%%%%%%%%%%%%%%%%
\vspace{3ex}                        % do not modify
%%%%%%%%%%%%%%%%%%%%%%%%%%%%%%%%%%%%%%%%%%%%%%%%%%%%%%%%%%%%%%%%%%%%%%%%
%% Authors
%%
\underline{V. Malafaia}$^{1}$\\
 M. T. Pe\~na$^{1,2}$\\
 J. Adam, Jr.$^{3}$ \\
%%%%%%%%%%%%%%%%%%%%%%%%%%%%%%%%%%%%%%%%%%%%%%%%%%%%%%%%%%%%%%%%%%%%%%%%
\vspace{2ex}                        % do not modify
%%%%%%%%%%%%%%%%%%%%%%%%%%%%%%%%%%%%%%%%%%%%%%%%%%%%%%%%%%%%%%%%%%%%%%%%
%%
\small{ $^{1}$ Centre for Theoretical Particle Physics, Instituto
Superior T\'ecnico , Av. Rovisco Pais, 1049-001 Lisbon, Portugal \\
$^{2}$ Department of Physics, Instituto Superior T\'ecnico ,
Av. Rovisco Pais, 1049-001 Lisbon, Portugal  \\
$^{3} $Nuclear Physics Institute, \v{R}e\v{z} near Prague,
    CZ-25068, Czech Republic }
%%%%%%%%%%%%%%%%%%%%%%%%%%%%%%%%%%%%%%%%%%%%%%%%%%%%%%%%%%%%%%%%%%%%%%%%
\end{center}                        % do not modify
\vspace{1ex}                        % do not modify
%%%%%%%%%%%%%%%%%%%%%%%%%%%%%%%%%%%%%%%%%%%%%%%%%%%%%%%%%%%%%%%%%%%%%%%%
%%                              PROCEEDING
%%    The proceednig for Invited talk should not exceed 8 pages,
%%    for contributed talks should not exceed 4 pages including title,
%%    author's, name(s), and institute(s), when processed by this macro.
%%
\normalsize{                        % do not modify

%% here paste Main Body  
% \section*{The S-matrix approach}  
% 
 Retardation effects and off-shell dependencies of the  rescattering
operator in neutral pion production were recently investigated\cite{MAP}
within the framework of time-ordered perturbation theory (TOPT). Various  
approximations to the result obtained from TOPT were   analysed. The central
problem underlying this study is that the final- and initial-state
interaction diagrams do not define a {\em single} effective operator. Since
in the time-ordered diagrams energy is not conserved at individual vertices,
each of these diagrams defines a different off-energy shell extension of the
pion re-scattering amplitude. This problem is not present in the S-matrix
derivation of the effective rescattering operator (for definition of the 
S-matrix approach and detailed discussion see Ref.~\cite{MAP} and references
therein).

In this work, we applied this approach to charged and neutral pion production
in $pp$ collisions, in which the spin/isospin channels filter
differently the rescattering mechanism. As in Ref.\cite{MAP}, the chiral
perturbation theory rescattering amplitude\cite{PMK} is employed. The
contributions from the single-nucleon (so-called, impulse approximation: IA),
$Z$-diagrams and $\Delta$-isobar mechanisms are also included.

%\section*{Results}
 The contributions of all the mechanisms considered are
shown on Fig.~\ref{tcrossd}. For the reaction $pp \rightarrow pp
\pi^{0}$ the IA term (dashed line) is suppressed. On the
other hand, the IA and rescattering mechanisms (dotted line),
which interfere destructively, are clearly not enough to describe
the data, and the $Z$-diagrams (dashed-dotted line) are found to
be very important in reproducing the cross section. The
$\Delta$-isobar, when included explicitly (solid line) plays a
significant role, even at energies close to threshold, improving
the description of the data.

\begin{table}
\vspace*{-0.5truecm}
\begin{center}
\begin{tabular}{ccccccccc}
\hline \multicolumn{1}{|c}{$\left( NN\right) _{i}$} &
%\multicolumn{1}{|c}{$\left( NN\right)_{f}\text{ }l^{\prime }$} &
\multicolumn{1}{|c}{$\left(NN\right)_{f}l^{\prime}$} &
\multicolumn{1}{|c}{$S$} & $L$ & $J$ &
\multicolumn{1}{|c}{$S^{\prime }$} & $L^{\prime }$ & $j^{\prime }$
& \multicolumn{1}{c|}{$l^{\prime }$} \\ \hline
\multicolumn{1}{|c}{$^{3}P_{0}$} & \multicolumn{1}{|c}{$\left(
^{1}S_{0}\right) s$} & \multicolumn{1}{|c}{$1$} & $1$ & $0$ &
\multicolumn{1}{|c}{$0$} & $0$ & $0$ & \multicolumn{1}{c|}{$0$} \\
\multicolumn{1}{|c}{$^{1}S_{0}$} & \multicolumn{1}{|c}{$\left(
^{3}P_{0}\right) s$} & \multicolumn{1}{|c}{$0$} & $0$ & $0$ &
\multicolumn{1}{|c}{$1$} & $1$ & $0$ & \multicolumn{1}{c|}{$0$} \\
\multicolumn{1}{|c}{$^{3}P_{0}$} & \multicolumn{1}{|c}{$\left(
^{3}P_{1}\right) p$} & \multicolumn{1}{|c}{$1$} & $1$ & $0$ &
\multicolumn{1}{|c}{$1$} & $1$ & $1$ & \multicolumn{1}{c|}{$1$} \\
\hline \multicolumn{1}{|c}{$^{3}P_{1}$} &
\multicolumn{1}{|c}{$\left( ^{3}P_{0}\right) p$} &
\multicolumn{1}{|c}{$1$} & $1$ & $1$ &
\multicolumn{1}{|c}{$1$} & $1$ & $0$ & \multicolumn{1}{c|}{$1$} \\
\multicolumn{1}{|c}{$^{3}P_{1}$} & \multicolumn{1}{|c}{$\left(
^{3}P_{2}\right) p$} & \multicolumn{1}{|c}{$1$} & $1$ & $1$ &
\multicolumn{1}{|c}{$1$} & $1$ & $2$ & \multicolumn{1}{c|}{$1$} \\
\multicolumn{1}{|c}{$^{3}P_{1}$} & \multicolumn{1}{|c}{$\left(
^{3}P_{1}\right) p$} & \multicolumn{1}{|c}{$1$} & $1$ & $1$ &
\multicolumn{1}{|c}{$1$} & $1$ & $1$ & \multicolumn{1}{c|}{$1$} \\
\hline \multicolumn{1}{|c}{$^{3}P_{2}$} &
\multicolumn{1}{|c}{$\left( ^{3}P_{1}\right) p$} &
\multicolumn{1}{|c}{$1$} & $1$ & $2$ &
\multicolumn{1}{|c}{$1$} & $1$ & $1$ & \multicolumn{1}{c|}{$1$} \\
\multicolumn{1}{|c}{$^{3}F_{2}$} & \multicolumn{1}{|c}{$\left(
^{3}P_{1}\right) p$} & \multicolumn{1}{|c}{$1$} & $3$ & $2$ &
\multicolumn{1}{|c}{$1$} & $1$ & $1$ & \multicolumn{1}{c|}{$1$} \\
\multicolumn{1}{|c}{$^{1}D_{2}$} & \multicolumn{1}{|c}{$\left(
^{3}P_{2}\right) s$} & \multicolumn{1}{|c}{$0$} & $2$ & $2$ &
\multicolumn{1}{|c}{$1$} & $1$ & $2$ & \multicolumn{1}{c|}{$0$} \\
\multicolumn{1}{|c}{$^{3}P_{2}$} & \multicolumn{1}{|c}{$\left(
^{3}P_{2}\right) p$} & \multicolumn{1}{|c}{$1$} & $1$ & $2$ &
\multicolumn{1}{|c}{$1$} & $1$ & $2$ & \multicolumn{1}{c|}{$1$} \\
\multicolumn{1}{|c}{$^{3}F_{2}$} & \multicolumn{1}{|c}{$\left(
^{3}P_{2}\right) p$} & \multicolumn{1}{|c}{$1$} & $3$ & $2$ &
\multicolumn{1}{|c}{$1$} & $1$ & $2$ & \multicolumn{1}{c|}{$1$} \\
\hline \multicolumn{1}{|c}{$^{3}F_{3}$} &
\multicolumn{1}{|c}{$\left( ^{3}P_{2}\right) p$} &
\multicolumn{1}{|c}{$1$} & $3$ & $3$ & \multicolumn{1}{|c}{$1$} &
$1$ & $2$ & \multicolumn{1}{c|}{$1$} \\ \hline
\end{tabular}%
\end{center}
\caption{The lowest partial waves for $pp \rightarrow pp \pi^{0}$.
$S(S')$, $L(L')$ and $J(j')$ are  the  spin, the
orbital momentum and the total angular momentum of the
initial(final) $NN$ pair, respectively. $l'$ is the  angular
momentum of the pion relative to the final $NN$ pair.  }
\label{pwpi0}
\end{table}

\begin{figure}
\vspace*{-1.5cm}
\begin{center}
\includegraphics[width=.75\textwidth,keepaspectratio]{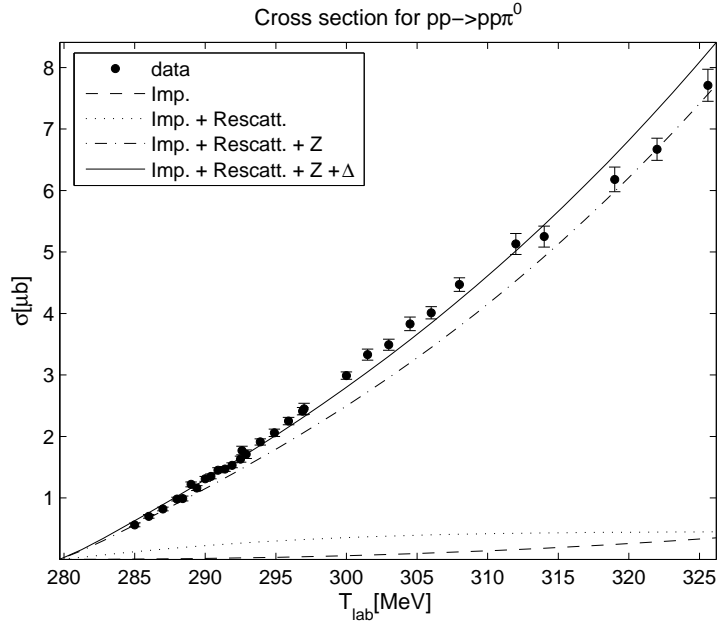}
\caption{Various contributions to the cross
section of $pp \rightarrow pp \pi^{0}$. The $NN$ interaction is
described by the Bonn B potential. The data points are from
Ref.\cite{HOM}} \label{tcrossd}
\end{center}
\vspace*{-0.5cm}
\end{figure}
\begin{figure}
\begin{center}
\includegraphics[width=.75\textwidth,keepaspectratio]{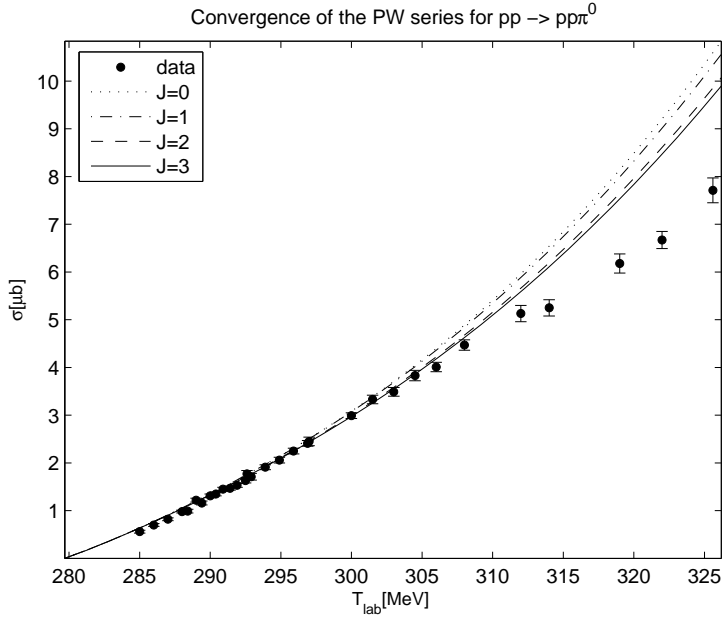}
\caption{Cross section for $pp \rightarrow pp \pi^{0}$. The
dotted, dashed-dotted, dashed and solid line correspond to the 
cross section for the Bonn B NN potential  and to all  
contributions up to $J=0, 1, 2, 3$, respectively.}
\label{pwtotaldcp}
\end{center}
\end{figure}
\begin{figure}
\begin{center}
\includegraphics[width=.75\textwidth,keepaspectratio]{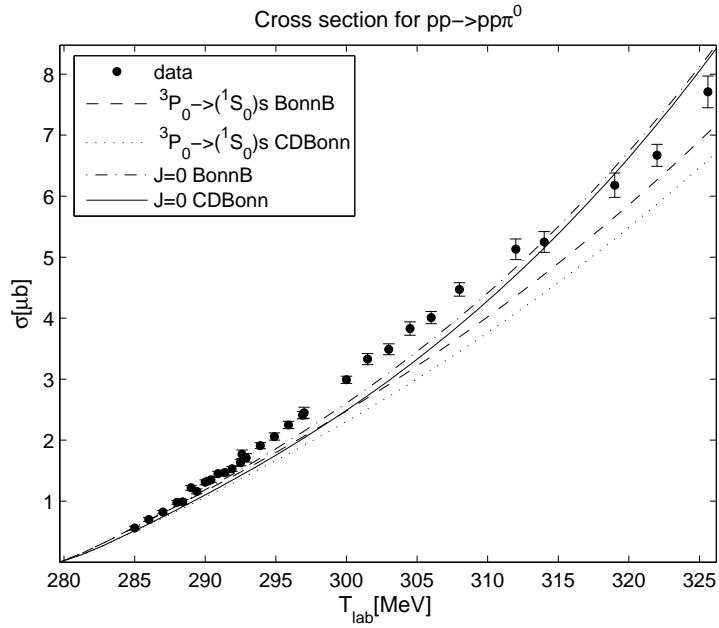}
\caption{Effect of the $NN$ interaction on the convergence of the
partial wave series for $pp \rightarrow pp \pi^{0}$. The
dotted(dashed) line corresponds to the cross section for the first
channel considered, $^{3}P_{0} \rightarrow \left( ^{1}S_{0}
\right) s$ for the Bonn B (CD Bonn) potential. The
solid(dashed-dotted) line corresponds to the contribution up to
$J=0$ for the Bonn B (CD Bonn) potential. The $\Delta$-isobar is
not included.} \label{compareCDB}
\end{center}
\end{figure}
\begin{figure} 
\begin{center}
\includegraphics[width=.75\textwidth,keepaspectratio]{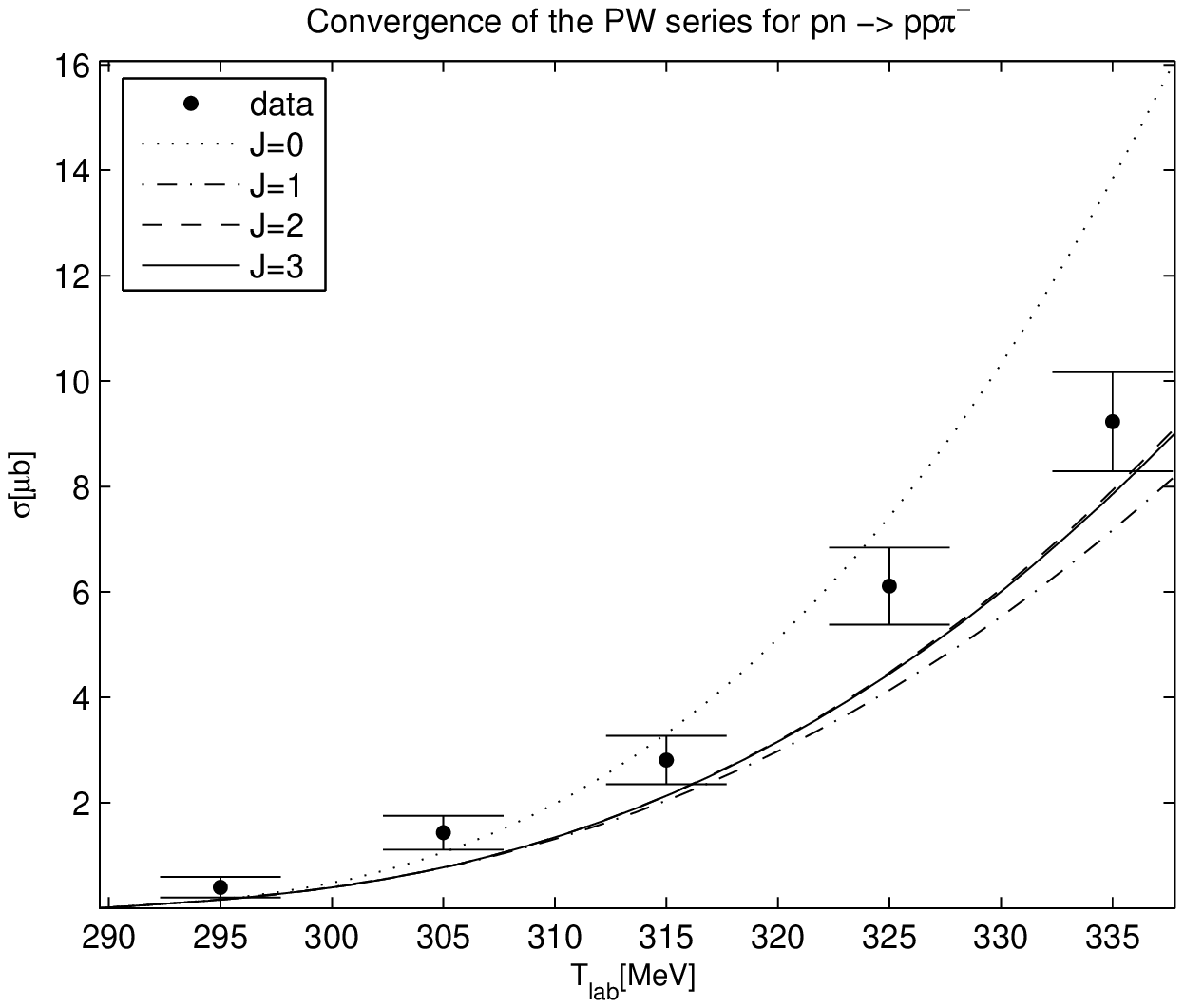}
\caption{The same of Fig.~\ref{pwtotaldcp}, but for $pn
\rightarrow pp \pi^{-}$. The data points are from
Ref.~\cite{Bachman:1995gn}.} \label{pwtotalmnd28}
\end{center}
\vspace*{-0.5truecm}
\end{figure}

The  convergence of the partial wave series of Table~\ref{pwpi0} is shown on
Fig.~\ref{pwtotaldcp}. As expected, the contributions with $J>0$ start to
play their role only  about 20MeV above threshold. Including all
contributions with $J$ up to $J=3$ yields a reasonably converged
results. The effect of the potential used for distorting the $NN$ final
and initial states is studied on Fig.~\ref{compareCDB}. Although the cross
sections for the first channel considered differ for the Bonn B and CD Bonn
potential (dotted and dashed-line, respectively), and this deviation is
increasing with $T_{lab}$, when all the contributions to $J=0$ are
included (dashed-dotted vs. solid line), the results for the cross section
for both potentials are very close.

For the reaction $pn \rightarrow pp \pi^{-}$, the convergence rate 
of the partial wave series is not as uniform as for $pp
\rightarrow pp \pi^{0}$ (Fig.~\ref{pwtotalmnd28}). Although $J=0$
is enough to describe the data close to threshold, the
contributions with higher $J$ included underestimate the cross
section, in particular for larger $T_{lab}$. The origin
of this discrepancy may be the non-inclusion of the coupled $N
\Delta$ channels, which play for sure an important role and must 
be included in a further analysis.

%\section*{Conclusions}

To sum it up: the S-matrix approach is successful in describing the cross
section for $\pi^{0}$ and $\pi^{-}$  production. For both reactions, the
convergence of the partial wave series is quite satisfactory and $J=0$ is
enough to describe the data close to threshold. If one includes all  
contributions with $J=0$  (not just a single dominant channel), the results
do not depend much on the choice of the $NN$ interaction. For charged pion
production, in which the $\Delta$ is known to play a decisive role (in
particular for $\pi^{+}$), a complete coupled-channel $N \Delta$ calculation
will be needed to describe the data. Also, the S-matrix approach should be
tested by more sensitive polarisation observables. This work was supported by
by FCT under Grant SFRD/BD/4876/2001 (V.M), 
POCTI/FNU/45831/2002 (M.T.P) and GA CR 202/03/0210 (J.A.).

 }
%%%%%%%%%%%%%%%%%%%%%%%%%%%%%%%%%%%%%%%%%%%%%%%%%%%%%%%%%%%%%%%%%%%%%%%%
%% References - leave commented if not needed

                   % do not modify
%%%%%%%%%%%%%%%%%%%%%%%%%%%%%%%%%%%%%%%%%%%%%%%%%%%%%%%%%%%%%%%%%%%%%%%%
\vspace{2ex}                        % do not modify
%% E-Mail of the presenting author
{\small \sl Contact e-mail: malafaia@cftp.ist.utl.pt}\\
%% Web page of the collaboration - please comment it if not needed
%{\small \sl Web page: http://.....}\\
%%
%%%%%%%%%%%%%%%%%%%%%%%%%%%%%%%%%%%%%%%%%%%%%%%%%%%%%%%%%%%%%%%%%%%%%%%%

\vspace{0.5cm}
\begin{thebibliography}{77}             % do not modify
\setlength{\itemsep}{0mm}                       % modifies line-spacing
\small                              % do not modify
\bibitem{MAP} V.~Malafaia, J.~Adam and M.~T.~Pena,
  Phys.\ Rev.\ C {\bf 71}, 034002 (2005).
\bibitem{PMK} B.-Y.~Park, {\it et al.}, Phys.~Rev.~{\bf C53}, 1519 (1996).
\bibitem{HOM} H.~O.~Meyer {\it et al.},  Nucl.\ Phys.\ A {\bf 539}, 633 (1992).
\bibitem{Bachman:1995gn}  M.~G.~Bachman {\it et al.},  Phys.\ Rev.\ C {\bf 52}, 495 (1995).
\end{thebibliography}
\end{document}